\documentclass[two column]{jpsj3}
\usepackage{txfonts}

\usepackage{color}
\usepackage{ulem}

\usepackage{bm}
\usepackage{braket}

\newcommand{\beginsupplement}{%
\setcounter{table}{0}
\renewcommand{\thetable}{S\arabic{table}}%
\setcounter{figure}{0}
\renewcommand{\thefigure}{S\arabic{figure}}%
}

\title{Possible Magnetic Structure with a Tilted Helical Plane in SmBe$_{13}$ Probed by $^9$Be-NMR Study}

\author{Hiroyuki Hidaka$^1$\thanks{hidaka@phys.sci.hokudai.ac.jp}, Yoshihiko Ihara$^1$, Tatsuya Yanagisawa$^1$, and Hiroshi Amitsuka$^1$}

\inst{$^1$Graduate School of Science, Hokkaido University, Sapporo, Hokkaido 060-0810, Japan \\
} 

\abst{
$^9$Be-NMR measurements were performed using single crystalline SmBe$_{13}$ in order to investigate a magnetic structure of a low-temperature ordering state microscopically. 
We observed a spectral broadening in the ordered state, and the broadened spectral shape depends on the magnetic field directions. 
By comparing the experimentally obtained and simulated NMR spectra for magnetic fields along the cubic [001] and [011] directions, we argue a helical structure with a basal plane tilted from the (001) plane in low-field ordering region under the assumption of a helical with a propagation vector of (0, 0, 1/3) found in other RBe$_{13}$ compounds (R = rare earths). 
Such a tilted helical structure is explained by a combination of an ellipse helical in the (001) plane and a longitudinal magnetic density wave along the [001] direction. 
Considering a magnetic easy axis parallel to [001] revealed by the magnetization measurements, the peculiar helical structure in SmBe$_{13}$ may be built on delicate balance among exchange interactions, dipole-dipole interaction, and single-ion magnetic anisotropy due to the crystalline electric field. 
}

\begin{document}
\maketitle


In magnetic materials, a competition among magnetic interactions, such as an exchange interaction between magnetic ions and the Zeeman interaction, leads to various types of magnetic structures. 
Among them, a helical ordering has been recently attracted much attention as a possible host magnetic structure for the emergence of an exotic spin textures, such as a magnetic skyrmion \cite{skyrmion1, skyrmion2, skyrmion3} and a chiral soliton \cite{YbNi3Al9, CrNb3S6}. 
The helical ordering can be classified into two types: a symmetric helical attributed to the competition between the exchange interactions, and a chiral helical attributed to the competition between ferromagnetic and Dzyaloshinskii--Moriya (DM) interactions. 
In addition to the above interactions, the introduction of the Zeeman interaction by external magnetic fields $H$ will induce a change in the magnetic structure, for instance conical or skyrmion ones. 
To deepen our understanding of the novel magnetic structures based on the helical ordering with multiple competing interactions, we should investigate characteristics of the helical state and its variation by magnetic fields from a microscopic point of view.

The beryllides RBe$_{13}$ (R = rare earths) are regarded as a typical system for the symmetric helical ordering, because of its rather simple crystal structure and well-localized 4$f$ electronic state \cite{Bucher, McElfresh, Hidaka-XRD}. 
These compounds crystallize in a NaZn$_{13}$-type face-centered-cubic structure with the space group $F$$m$$\bar{\rm 3}$$c$ (No. 226, $O_h^{\rm 6}$) \cite{McElfresh}, as shown in Fig. 1(a). 
The R ions, occupying the 8$a$ site with the site symmetry $O$, form a simple cubic. 
In this crystal structure, the DM interaction is absent. 
It is also known that many RBe$_{13}$ compounds with R = Gd--Er undergo a proper helical ordering formed by well-localized 4$f$ electrons of the R$^{3+}$ ions, whose propagation vector \mbox{\boldmath $Q$} is commonly $\sim$ (0, 0, 1/3) \cite{Vigneron}. 
This helical structure has been explained by a competition between intralayer and interlayer exchange interactions for a one-dimensional layer crystal \cite{Becker}. 
However, the magnetic structures in the light-rare-earth RBe$_{13}$ compounds, such as NdBe$_{13}$ and SmBe$_{13}$, have not been experimentally determined thus far, and it is unclear whether the same model as for the heavy-rare-earth systems can be applicable.

\begin{figure}[tb]
\begin{centering}
\includegraphics[width=0.45\textwidth]{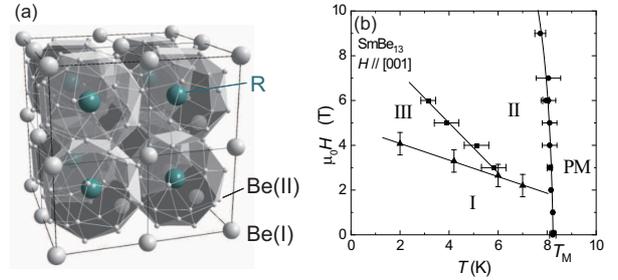}
\caption{
(Color online) (a) Crystal structure of RBe$_{13}$ and (b) the $H$ -- $T$ phase diagram for \mbox{\boldmath $H$} $\parallel$ [001] in SmBe$_{13}$ constructed in the previous study \cite{Hidaka-SmBe13}. 
}
\label{Fig1}
\end{centering}
\end{figure}

In the present study, we focus our attention on the magnetic structure of SmBe$_{13}$, which shows a magnetic ordering of the localized
4$f$ moments of Sm$^{3+}$ ions at $T_{\rm M}$ $\sim$ 8.3 K \cite{Hidaka-SmBe13, Tsutsui}. 
The previous studies on polycrystalline samples suggested a crystalline-electric-field (CEF) level scheme with a $\Gamma_7$ doublet ground state and a $\Gamma_8$ quartet first-excited state (the $\Gamma_7$--$\Gamma_8$ level scheme) \cite{Bucher, Besnus}. 
On the other hand, our recent studies on single crystals revealed that the CEF level scheme is the $\Gamma_8$ quartet ground state and the $\Gamma_7$ doublet first-excited state with energy separation of 90 K (the $\Gamma_8$--$\Gamma_7$ level scheme) \cite{Hidaka-SmBe13, Mombetsu}. 
In addition, its magnetic field--temperature ($H$ -- $T$) phase diagram for \mbox{\boldmath $H$} $\parallel$ [001] consists of three regions below $T_{\rm M}$, as shown in Fig. 1(b). \cite{Hidaka-SmBe13, Mombetsu}. 
To determine the magnetic structures in these ordering regions microscopically, neutron scattering is one of the most powerful methods, but there is a difficulty that a $^{149}$Sm nucleus is a neutron absorber. 
Therefore, we performed the nuclear magnetic resonance (NMR) measurements on single crystalline sample in this study. 
From the obtained results of $^9$Be-NMR spectra, we propose that the magnetic structure of SmBe$_{13}$ is a helical with a tilted basal plane rather than the proper helical. 
The magnetic structure of the high-field region will also be discussed.


Single crystals of SmBe$_{13}$ were grown by the Al-flux method, as described in the previous report \cite{Hidaka-SmBe13}. 
The $^9$Be-NMR measurements were performed by conventional spin-echo method using a single crystal assigned as sample 1 for magnetic fields applied parallel to the [001] and [011] directions. 
The size of the sample 1 used for the NMR measurements was $\sim$ 3 $\times$ 3 $\times$ 1 mm$^3$. 
In the present crystal structure, Be occupies two crystallographically independent 8$b$ site [Be(I)] and 96$i$ site [Be(II)]. 
The NMR spectrum mainly originates from the Be(II) sites because of a larger number of nuclei in the unit cell, Be(I): Be(II) = 1:12.  
The details of analyses of the obtained spectrum are indicated in the Supplemental Materials (SM) \cite{SM}. 
The DC magnetization ($M$) measurements were also performed for investigating the magnetic anisotropy using sample 2 in magnetic fields up to 7 T by a magnetic property measurement system (MPMS, Quantum Design Inc.). 
The magnetic field was applied along the cubic [001], [011], and [111] directions. 
The weight of the sample 2 used for the $M$ measurements was $\sim$ 10 mg.


\begin{figure}[tb]
\begin{centering}
\includegraphics[width=0.35\textwidth]{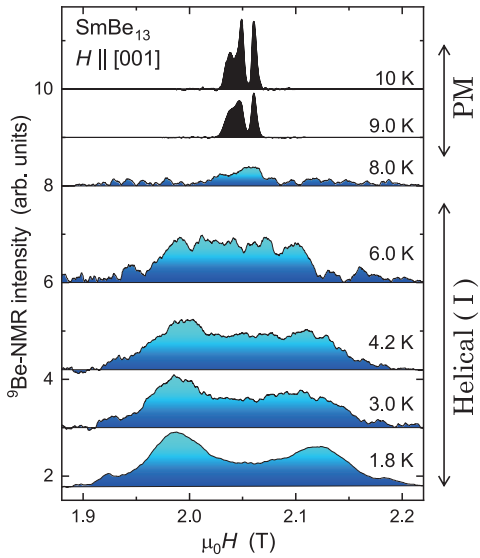}
\caption{
(Color online) The field-sweep $^9$Be-NMR spectra obtained at a fixed frequency of $12.31$ MHz. The magnetic field was applied along the [001] direction.
The NMR intensity once becomes weak near the magnetic ordering temperature $T_{\rm M} \sim 8$ K because of the critical fluctuations and recovers in the ordered state with significant spectral broadening.
The three-peaks structure above $T_{\rm M}$ is explained by the electric quadrupole splitting with highly anisotropic NMR shift. 
}
\label{Fig2} 
\end{centering}
\end{figure}

Figure 2 exhibits $^9$Be-NMR spectra of SmBe$_{13}$ obtained between 1.8 and 10 K by sweeping the magnetic fields near 2 T at a fixed frequency $f_0 = 12.31$ MHz. 
The field direction is parallel to the [001] axis. 
In the paramagnetic (PM) state at 9 and 10 K, the relatively sharp resonance peaks were observed near the resonance magnetic field $\mu_{\rm 0} H_0 = f_0/\gamma = 2.06$ T. 
Here, $\gamma = 5.9833$ MHz/T is the gyromagnetic ratio for $^{9}$Be nuclear moment. 
Below $T_{\rm M}$ $\sim$ 8 K, we observed a significant spectrum broadening associated with the magnetic phase transition. 
In the $H$ -- $T$ phase diagram constructed in the previous bulk measurements \cite{Hidaka-SmBe13}, this magnetically ordered state corresponds to the low-field one, named as region I. 
The broad linewidth of 300 mT in region I completely covers the three-peak structures separated by 30 mT constructed by the nuclear electric quadrupole interaction (NQR splitting, see SM). \cite{SM}
We, therefore, assume a simplified Gaussian broadening for the spectrum analyses in the ordered state. 
The internal fields at the Be(II) site are dominated by the isotropic transferred hyperfine coupling $A_{\rm iso}$ between neighboring two Sm moments, which was determined from the slope in the $K$-$\chi$ plot \cite{SM}. By equally dividing the total $A_{\rm iso}$ to two Sm moments, we estimated the coupling constant for one neighboring Sm moment as 0.13 T/$\mu_{\rm B}$. 
The anisotropic part of the hyperfine coupling constant was also estimated, and the obtained  $\bm{A}_{\rm aniso}$ = (0.07, -0.05, 0.02) T/$\mu_{\rm B}$ was ascribed to the dipole contribution.\cite{SM}

\begin{figure}[tb]
\begin{centering}
\includegraphics[width=0.35\textwidth]{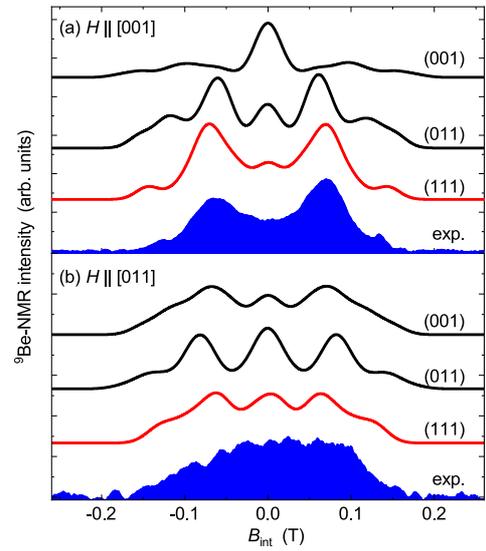}
\caption{
(Color online) 
The $^9$Be-NMR spectra in magnetic fields near 2 T applied parallel to the (a) [001] and (b) [011] directions. 
The experimentally obtained spectra at 1.8 K (fill color) are compared with the simulation for three different helical structures (solid lines). 
We suggest that the tilted helical with the (111) basal plane can explain the experimental results for both field directions, namely a two-peak structure for [001] and a trapezoidal structure for [011]. 
}
\label{Fig3}
\end{centering}
\end{figure}

\begin{figure}[tb]
\begin{centering}
\includegraphics[width=0.45\textwidth]{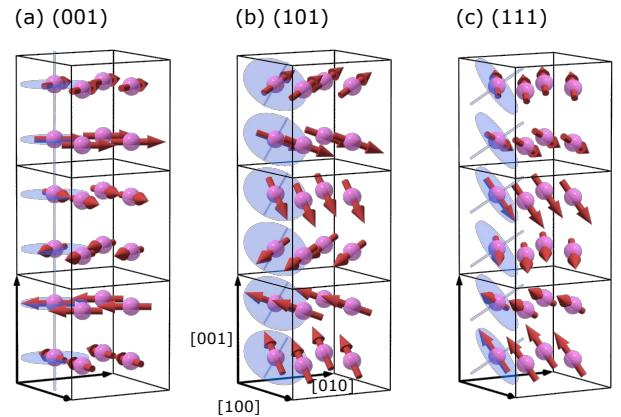}
\caption{(Color online) (a)-(c) Magnetic structures with three different helical planes for (001), (101), and (111), respectively. 
Here, the magnetic structure for (101) is shown instead of (011) for clarity. 
}
\label{Fig4}
\end{centering}
\end{figure}

In the region I, we measured $^9$Be-NMR spectra for two different magnetic field orientations. 
The spectra for \mbox{\boldmath $H$} $\parallel$ [001] and for \mbox{\boldmath $H$} $\parallel$ [011] are shown in Figs. 3(a) and 3(b), respectively. 
The horizontal axis of Fig. 3 is defined as $B_{\rm int} = f_0/\gamma-\mu_0H$, which measures the internal fields at the Be sites projected to the external field direction. 
We found a trapezoidal spectrum for \mbox{\boldmath $H$} $\parallel$ [011], while two peaks at both edges were observed for \mbox{\boldmath $H$} $\parallel$ [001]. 
To explain these spectral shapes, we simulated the NMR spectra for several helical magnetic structures. 
In the present analyses, we assume that the propagation vector \mbox{\boldmath $Q$} is fixed to (0, 0, 1/3), which has been commonly found in the RBe$_{13}$ systems showing the helical ordering. 
In a cubic crystal structure, six magnetic propagation vectors $\bm{Q}$ of $(0,0,\pm1/3), (0,\pm1/3,0)$ and $(\pm1/3,0,0)$ are equivalent. 
Thus, to simulate the NMR spectrum, we assume that the six domains are equally distributed. 
When the helical plane is (001), the magnetic moments of Sm at a position \mbox{\boldmath $r$}$_i$ are written as 
\begin{equation} 
\bm{M}_{\rm (001)} = {\it M}_{\rm 0} \left( \cos (\mbox{\boldmath $Q$} \cdot \mbox{\boldmath $r$} + \phi_0), \sin (\mbox{\boldmath $Q$} \cdot \mbox{\boldmath $r$} + \phi_0), 0 \right).
\end{equation} 
Here, $M_{\rm 0}$ and $\phi_{\rm 0}$ are the size of Sm magnetic moment ($\sim$ 0.5 $\mu_{\rm B}$) and arbitrary initial phase, respectively. 
We also modeled several other helical structures, which possess the same \mbox{\boldmath $Q$}, but tilted helical plane. 
In the case of helical plane parallel to (011) [Fig. 4(b)] and (111) [Fig. 4(c)], the directions of Sm moments are written as
\begin{equation} 
\begin{split} 
\bm{M}_{\rm (011)} = {\it M}_{\rm 0} \bigl( &\cos (\mbox{\boldmath $Q$} \cdot \mbox{\boldmath $r$} + \phi_0), \\ 
&\sin (\mbox{\boldmath $Q$} \cdot \mbox{\boldmath $r$} + \phi_0)/\sqrt{2}, \\
&\sin (\mbox{\boldmath $Q$} \cdot \mbox{\boldmath $r$} + \phi_0)/\sqrt{2} \bigr). 
\end{split}
\end{equation} 
\begin{equation} 
\begin{split} 
\bm{M}_{\rm (111)} = {\it M}_{\rm 0} \bigl( &\cos (\mbox{\boldmath $Q$} \cdot \mbox{\boldmath $r$} + \phi_0)/\sqrt{2} + \sin (\mbox{\boldmath $Q$} \cdot \mbox{\boldmath $r$} + \phi_0)/\sqrt{6}, \\ 
&-\cos (\mbox{\boldmath $Q$} \cdot \mbox{\boldmath $r$} + \phi_0)/\sqrt{2} + \sin (\mbox{\boldmath $Q$} \cdot \mbox{\boldmath $r$} + \phi_0)/\sqrt{6}, \\
&-2\sin (\mbox{\boldmath $Q$} \cdot \mbox{\boldmath $r$} + \phi_0)/\sqrt{6} \bigr). 
\end{split}
\end{equation} 
Here, we defined that the magnetic structures shown in Fig. 4 correspond to $\phi_0 = 0$, and we do not consider that the propagation vector is parallel to the helical plane.

The internal fields at the Be(II) site between Sm sites $i$ and $j$ are then estimated by the sum of the hyperfine fields from neighboring Sm moments and dipole fields as $\bm{B}_{\rm int} = A_{\rm iso}(\bm{M}_i+\bm{M}_j)+ \bm{B}_{\rm dip}$. 
The hyperfine fields give the predominant contribution for the peak splitting of approximately $\pm 100$ mT and the dipole fields of less than $20$ mT contribute to the broadening of each peak. 
The shift in NMR spectrum is caused by the internal-field component parallel to the external field direction. To take into account the randomly distributed domains, we calculated the internal fields along the six directions equivalent to cubic [001], and twelve directions equivalent to [011]. 
By including all these internal-field contributions, the $^9$Be-NMR spectrum in the ordered state is simulated as shown in Fig. 3. 
The two-peak spectrum for \mbox{\boldmath $H$} $\parallel$ [001] and the trapezoidal spectrum for \mbox{\boldmath $H$} $\parallel$ [011] is the most consistently explained by the (111) plane model with $\phi_0$ = 0, leading us to suggest a helical structure with a tilted basal plane for region I. 
The results of simulation for other $\phi_0$ are presented in SM. \cite{SM}

The tilted helical structure suggested here for region I of SmBe$_{13}$ has not been found in other RBe$_{13}$ compounds, which show a proper helical ordering \cite{Vigneron}. 
Here, we discuss the mechanism for the novel helical structure in SmBe$_{13}$. 
The helical structure in RBe$_{13}$ has been explained by a competition of the Heisenberg exchange interactions assuming a one-dimensional layer crystal \cite{Becker}. 
Based on this explanation, the mechanism of the proper helical ordering has been recently revealed in GdBe$_{13}$, which has no total-orbital momentum $L$ and thus no single-ion anisotropy \cite{Hidaka-GdBe13}. 
The propagation vector \mbox{\boldmath $Q$} is determined by the Ruderman-Kittel-Kasuya-Yosida (RKKY) interaction via anisotropic Fermi surfaces, and thus we can reasonably assume that SmBe$_{13}$ has the same \mbox{\boldmath $Q$} because well-localized 4$f$ electrons yield similar Fermi surfaces for all RBe$_{13}$ systems.  
In contrast, the dipole--dipole interaction between the 4$f$ moments orients the helical plane perpendicular to the direction of \mbox{\boldmath $Q$} in the isotropic GdBe$_{13}$, which contradicts to the tilted helical structure.

In addition to the interactions considered for GdBe$_{13}$, the single-ion anisotropy should be an important factor to construct a magnetic structure for a compound containing the R$^{3+}$ ions with non-zero $L$. 
Therefore, to reveal the magnetic anisotropy in SmBe$_{13}$, we measured the $M$ process at 12 and 2 K for three crystallographic axes of [001], [011], and [111], and the results are shown in Figs. 5(a) and 5(b), respectively. 
The results observed at 12 K indicate that the magnetization easy axis is the cubic [001] direction in the PM state. 
The easy axis determined experimentally is consistent with a model calculation considering the $\Gamma_8$--$\Gamma_7$ level scheme \cite{Hidaka-SmBe13}. 
Note that the CEF calculation considering the $\Gamma_7$--$\Gamma_8$ level scheme proposed by Besnus et al. \cite{Besnus} predicts that the magnetization easy axis is the [111] direction, which is inconsistent with the present experimental results. 
Details of the CEF calculations are shown in SM \cite{SM}.

\begin{figure}[tb]
\begin{centering}
\includegraphics[width=0.53\textwidth]{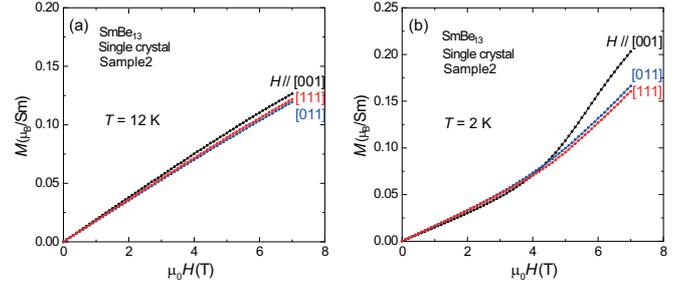}
\caption{(Color online) Magnetization processes of SmBe$_{13}$ at (a) 12 K and (b) 2 K for \mbox{\boldmath $H$} $\parallel$ [001], [011], and [111].}
\label{Fig5}
\end{centering}
\end{figure}

How can the tilted helical ordering occur in SmBe$_{13}$? 
The equation (3) indicates that the tilted helical structure with (111) basal plane is composed of an elliptical helical in the (001) plane and a longitudinal magnetic density wave along the [001] direction. 
Since the single-ion anisotropy attributed to the CEF effect aligns the magnetic moments along the easy axis, the magnetic component in the [001] direction may order, leading to the tilted helical structure in region I of SmBe$_{13}$. 
In addition, the hard axis parallel to the [111] direction, predicted from the CEF calculation, may be a reason for the (111) helical plane, which avoids the magnetic moments directed in the [111] direction. 
Thus, we argue that the competition among the Heisenberg exchange interactions due to the RKKY interaction dominates the formation of the helical structure and its \mbox{\boldmath $Q$} vector, and the single-ion anisotropy and the dipole-dipole interaction determine the direction of the magnetic moments, namely the orientation of the helical plane. 
It has been known that MnWO$_4$ also exhibits similar tilted helical structure, where the competition between isotropic exchange interaction and single-ion anisotropy has been suggested for its origin. \cite{MnWO4-Lautenschlager}
In the case of SmBe$_{13}$ with the $\Gamma_8$ ground state, we should also keep in mind that higher-order multipoles may be involved with the peculiar magnetic structure.

On the basis of the second-order phase transition in the Landau's theory, the tilted helical structure proposed in the present study cannot be described by a single irreducible representation, and thus symmetry reductions at least two times, including the transition at $T_{\rm M}$, are required. 
This consideration suggests that multiple magnetic structures can appear in the $H$ -- $T$ phase diagram even at zero magnetic field.  
On the other hand, one possibility for another magnetic structure is that with different \mbox{\boldmath $Q$}, which may also explain the observed NMR spectra. 
In this case, we need to find out the reason why only SmBe$_{13}$ is the exception in the RBe$_{13}$ family.

\begin{figure}[tb]
\begin{centering}
\includegraphics[width=0.42\textwidth]{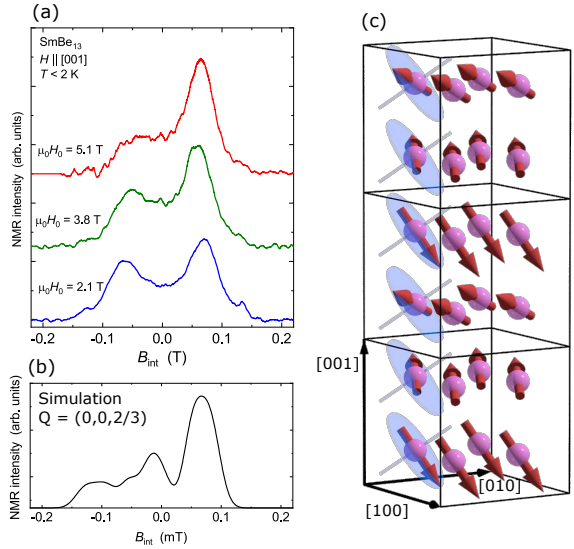}
\caption{(Color online) (a) The $^9$Be-NMR spectrum at magnetic fields around 5 T at 2 K, and (c) an up-up-down magnetic structure. The external field direction is parallel to [001]. The solid line shown in (b) represents the simulation result assuming the up-up-down structure.}
\label{Fig6}
\end{centering}
\end{figure}

Finally, we comment on a magnetic structure in higher-field region III of SmBe$_{13}$.
Figure 6(a) shows the $^9$Be-NMR spectra for \mbox{\boldmath $H$} $\parallel$ [001] measured around 5.1 and 3.8 T ($T$ = 2 K) and 2.1 T ($T=1.8$ K). 
The asymmetry of NMR spectrum progressively increases at high fields with higher intensity at larger internal fields, indicating that the magnetic structure in region III differs from that in region I. 
The overall spectrum width is not field dependent. 
A helical magnetic structure with $\bm{Q} = (0,0,1/3)$ cannot explain the asymmetric spectrum structure, because every one third Sm moments are always antiparallel to each other, which results in a symmetric distribution of internal fields. 
A small asymmetry that already exists at 2.1 T may be interpreted that the Sm moments partly form a structure for region III even below the magnetic field where the apparent kink appears in the magnetization curve as shown in Fig. 5(b). 
We need to implement unbalance in up-down spin population along the [001] direction, such as up-up-down arrangement with $\bm{Q}= (0, 0, 2/3)$, as displayed in Fig. 6(c). 
The internal fields for $\bm{Q}=(0,0,2/3)$ with tilted helical plane was calculated and the obtained spectrum is shown in Fig.~\ref{Fig6}(b). 
For this calculation, we take into account three magnetic domains with positive $Q$ components, assuming an alignment by external fields. 
The spectral shape at $5.1$ T is almost consistent with the simulated spectrum with some difference at the negative internal fields. 
We propose that the Sm moments are not uniformly distributed in the helical plane in high fields.

This magnetic structure is also motivated by the $M$($H$) curve for [001] at 2 K, which becomes apparently larger than that for [011] and [111] above $\sim$ 4 T [Fig. 5(b)], where the ordered state changes from region I to III \cite{Hidaka-SmBe13}. 
The change in the magnetic structure induced by $H$ has also been reported in HoBe$_{13}$, and it has been pointed out that the appearance of the high-field phase is provided by the single-ion anisotropy \cite{Dervenagas}. 
However, it is unclear whether the change in \mbox{\boldmath $Q$} from (0, 0, 1/3) to (0, 0, 2/3) actually occurs in SmBe$_{13}$, which may be interpreted as a change in the relative strength of the exchange interactions by applying high magnetic field. 
It is necessary to determine the magnetic structure of SmBe$_{13}$ for each magnetic region by other microscopic measurement methods, such as neutron scattering with isotope substitution and resonant X-ray scattering.


In summary, we performed $^9$Be-NMR measurements for \mbox{\boldmath $H$} $\parallel$ [001] and [011] and $M$ measurements for \mbox{\boldmath $H$} $\parallel$ [001], [011], and [111] using single crystals to investigate magnetic structures below $T_{\rm M}$ in SmBe$_{13}$. 
The observed NMR spectral shapes in low-field region I cannot be explained by a proper helical magnetic structure with \mbox{\boldmath $Q$} = (0, 0, 1/3), but rather suggests the possibility of a helical structure with a tilted basal plane parallel to (111). 
This tilted helical structure can be explained by a combination of an elliptical helical in the (001) plane and a longitudinal magnetic density wave along the [001] direction due to the single-ion magnetic anisotropy. 
Clarifying why such a tilted helical ordering occurs only in SmBe$_{13}$ among the RBe$_{13}$ family will help to deepen general understanding of the role of magnetic anisotropy in the helical ordering. 
In addition, we proposed an up-up-down type magnetic structure from the NMR spectrum in higher-field region III.

\begin{acknowledgment}

The authors are grateful to A. Koriki and C. Tabata for helpful discussions. 
The present research was supported by JSPS Grants-in-Aid for Scientific Research (KAKENHI) Grants Numbers JP20224015(S), JP25400346(C), JP26400342(C), JP19H01832(B), JP20K03825(C), JP15H05882(J-Physics), JP15H05885(J-Physics), and JP18H04297(J-Physics). 
This study was also partly supported by Hokkaido University, Global Facility Center (GFC), Advanced Physical Property Open Unit (APPOU), funded by MEXT under ``Support Program for Implementation of New Equipment Sharing System''.  

\end{acknowledgment}

\beginsupplement

\section{Estimation of hyperfine coupling constant}
To estimate the anisotropic hyperfine coupling constant of SmBe$_{13}$, we need to understand the $^9$Be-NMR spectrum. 
First we will assign the peaks in the Be-NMR spectrum to corresponding Be sites.
In the crystal structure of RBe$_{13}$, Be occupies two crystallographically independent 8$b$ site [Be(I)] and 96$i$ site [Be(II)]. 
The NMR spectrum mainly originates from Be(II) sites because of a larger number of nuclei in a unit cell, Be(I):Be(II) = 1:12. 
The signal from Be(I) site was not identified in this study. Since the electric field gradient (EFG) at Be(II) site is finite because of a low site symmetry, three NMR peaks from $m$ = 3/2 $\Leftrightarrow$ 1/2, 1/2 $\Leftrightarrow$ -1/2, and -1/2 $\Leftrightarrow$ -3/2 transitions will be observed for the $^9$Be nuclear spins with $I$ = 3/2. 
Besides, when a magnetic field is applied to [001] direction, the crystallographic three-fold symmetry is magnetically broken, and thus three sets of three-peaks will be observed. 
The magnetically nonequivalent three sets of peaks are labeled as R, G, and B as shown in Fig. S1(a). 
We clearly resolved NMR signals from these sites at 60 K because of rather small linewidth as shown in Fig. S1(b). 
The highly resolved spectrum at 60 K shows a good sample quality and confirms that the external field is aligned parallel to the [001] direction, because even a small misalignment further lowers the local symmetry which results in a significant line broadening. 
The three sets of Be(II) spectra are decomposed as indicated at the bottom of Fig. S1(b), and the NQR frequencies obtained by the separation of each set of peaks are $n_{\rm Q}$ = (17, 68, 85) kHz. 
Similar results were reported for a sister compound UBe$_{13}$, in which the direction for the smallest NQR frequency is perpendicular to the mirror plane, that is, parallel to the crystalline axis \cite{SM1}. 
The other principal axes of EFG in the mirror plane are slightly tilted from the crystalline axes. 
According to the result for UBe$_{13}$, we assign these spectra in SmBe$_{13}$ to the crystal structure as represented in Fig. S1(b).

\begin{figure}[tb]
\begin{centering}
\includegraphics[width=0.4\textwidth]{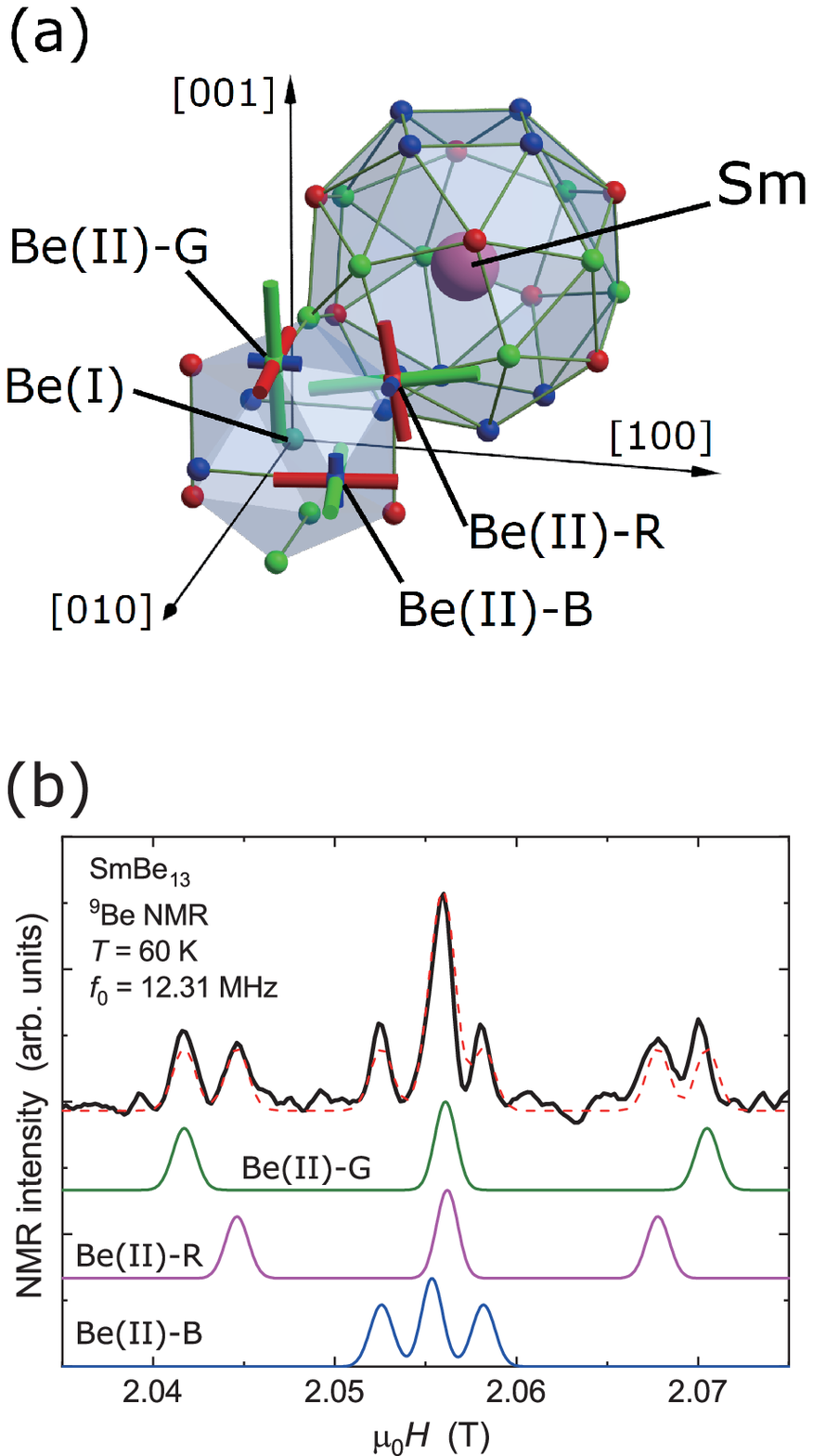}
\caption{
(Color online) (a) Be(II) sites assigned for three sets of $^9$Be NMR spectra. EFG principal axes are represented as bars on the Be sites. (b) $^9$Be NMR spectrum in the field applied parallel to [001] direction at 60 K. 
}
\label{Fig. S1}
\end{centering}
\end{figure}

Based on the spectrum assignment described above, we can measure the anisotropic Knight shift for three different field directions with respect to the EFG principal axes at the same time. 
The temperature dependence of the Knight shift $K$ for each site is shown in Fig. S2. 
The Knight shift was measured at approximately 5 T for better resolution. 
The temperature dependence of Knight shift scales to that of the bulk susceptibility $\chi$($T$) above 15 K for all sites. \cite{Hidaka-SmBe13} 
At lower temperatures the peak positions were not determined precisely because of the spectrum broadening near the magnetic phase transition, where the linewidth becomes comparable to the NQR splitting, as indicated in Fig. 2 of the main text. 
Within the temperature range between 100 and 15 K, a linear relationship between $K$ and $\chi$ was observed as demonstrated by the $K$--$\chi$ plot shown in the inset to Fig. S2. 
From the slope in the $K$--$\chi$ plot, we obtain the anisotropic hyperfine coupling constants as $A_{\rm hf}$ = 0.33, 0.21, 0.24 T/$\mu_{\rm B}$. 
This hyperfine coupling constants can be decomposed into symmetric and asymmetric terms as $A_{\rm hf}$ = 0.26 + (0.07, -0.05, -0.02) T/$\mu_{\rm B}$. 
The asymmetric term is explained purely by the direct dipole coupling from the localized Sm spins, which is calculated to be (0.061, -0.053, -0.016) T/$\mu_{\rm B}$ by summing up the dipole fields from Sm moment within 100 $\AA$ from the target Be(II) site. 
This is in contrast with the results for UBe$_{13}$, where a dipole coupling to the Be $p$ electrons also contribute to the asymmetric hyperfine fields because of the hybridization between Be $p$ electrons with U 5$f$ electrons \cite{SM1}. 
A negligibly small $p$-electron contribution in SmBe$_{13}$ suggests a weak $p$-$f$ hybridization.

\begin{figure}[tb]
\begin{centering}
\includegraphics[width=0.4\textwidth]{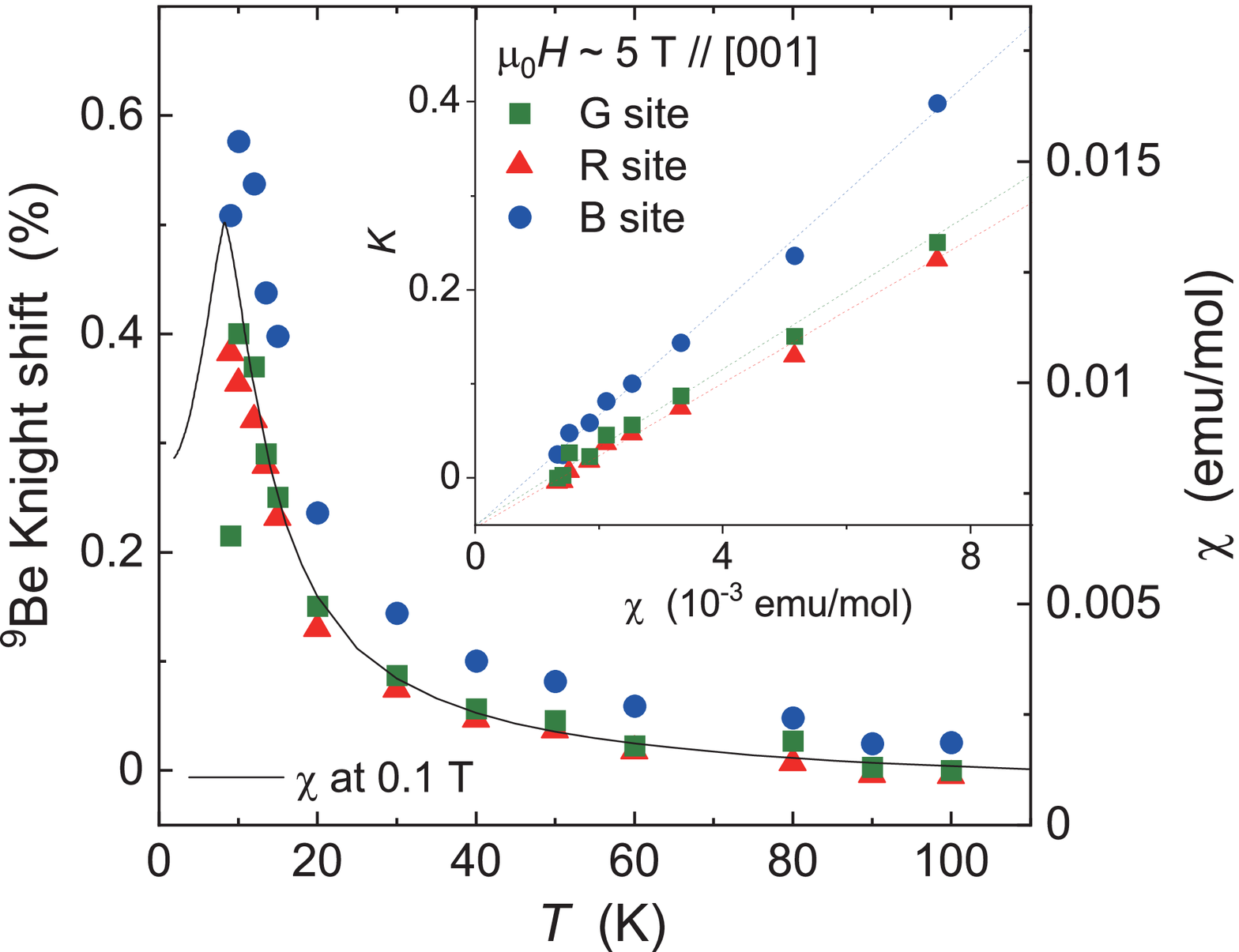}
\caption{
(Color online) Temperature dependence of Knight shift for three Be(II) sites. Inset shows the $K$--$\chi$ plot for each site. 
}
\label{FigS2}
\end{centering}
\end{figure}

\section{Initial phase dependence of NMR spectra}

\begin{figure}[tb]
\begin{centering}
\includegraphics[width=0.47\textwidth]{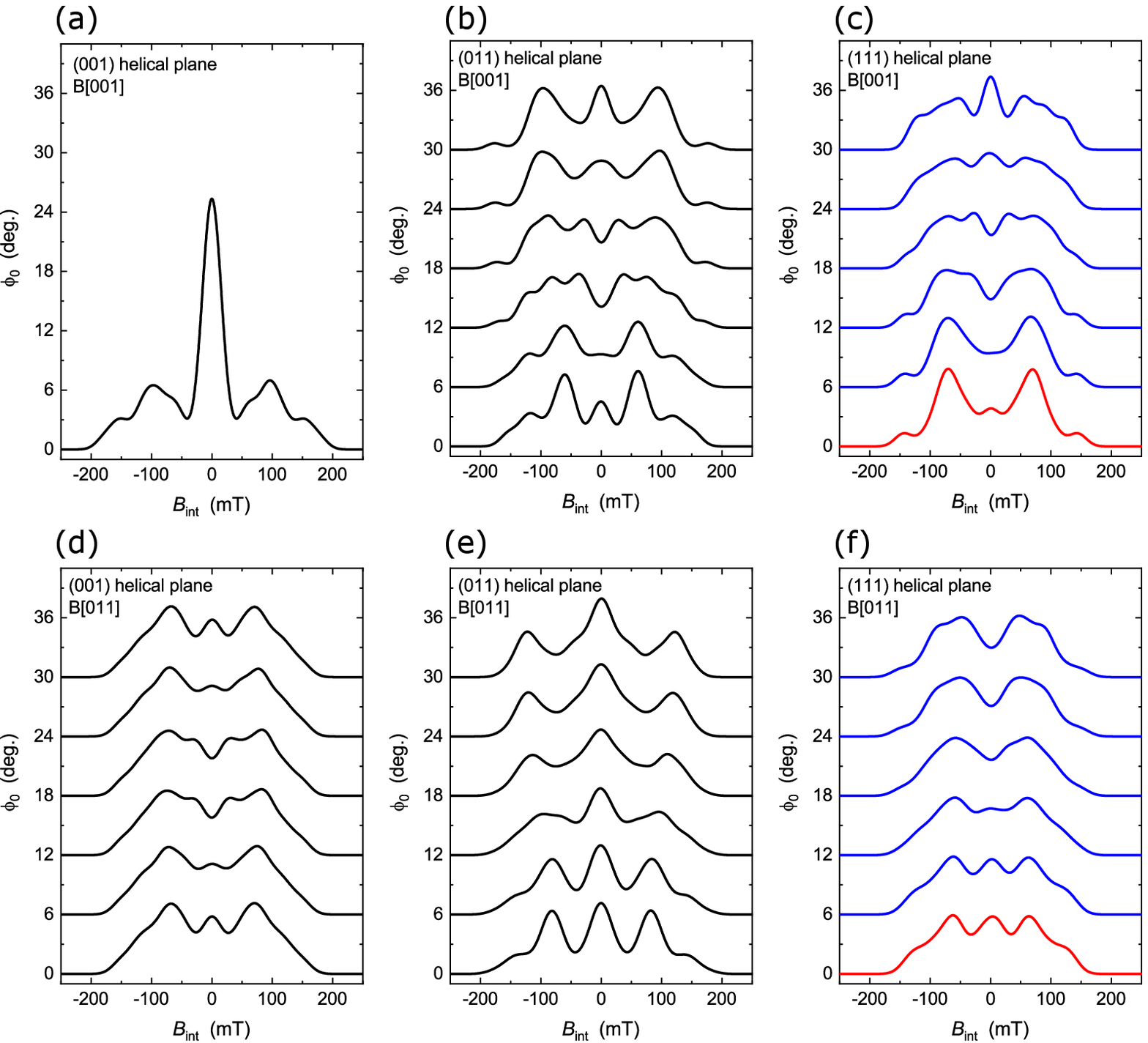}
\caption{
The $\phi_0$ dependence of NMR spectra assuming (a) the (001) helical plane, (b) the (011) helical plane, and (c) the (111) helical plane in magnetic fields applied along the [001] direction, and (d) the (001) helical plane, (e) the (011) helical plane, and (f) the (111) helical plane in magnetic fields applied along the [011] direction. 
Each spectrum is offset by $\phi_0$ for good visibility. 
Only one spectrum is shown in (a) because no $\phi_0$ dependence was found for (001) helical plane in [001] field direction. 
The two-peak structure in [001] field direction and trapezoidal structure in [011] field direction are obtained at $\phi=0$ in (111) helical plane. 
}
\label{FigS3}
\end{centering}
\end{figure}

In the helical magnetic structure, Sm moments are aligned in the helical plane. 
Such a structure has a freedom of global rotation within the helical plane, which is parameterized by the initial phase $\phi_0$ in equations (1)-(3) in the main text. 
The internal fields at the Be(II) sites change with $\phi_0$, and thus the NMR spectral shape has $\phi_0$ dependence. 
In contrast, the bulk magnetization does not depend on this microscopic parameter. 
We defined that the magnetic structures shown in Fig. 4 correspond to $\phi_0 = 0$. 
In Fig. S3, we show the spectral shapes for other $\phi_0$ calculated for three helical planes of (001), (011), and (111) and two field directions parallel to [001] and [011]. 
In the case of proper helical [Fig. S3(a)] one spectrum for $\phi_0=0$ is shown, as the internal field distribution along the external field direction does not depend on $\phi_0$. 
For other cases, we calculated the spectra for $0^{\circ} < \phi_0 < 30 ^{\circ}$ as the shift in $\phi_0$ by $60^{\circ}$ gives the identical spectral shape, 
reflecting the six-fold translational symmetry. 
Among all these spectra $\phi_0=0$ for (111) helical plane fit to the experimental results both in fields along [001] and [011] directions, which leads us to suggest the tilted helical structure as we discussed in the main text.

\section{Magnetization based on the CEF model}

We show the magnetization of SmBe$_{13}$ calculated based on the crystalline-electric-field (CEF) effect. 
The magnetization can be calculated by 
\begin{equation} 
M = g_J\mu_{\rm B} \sum_{n} \frac{\bra{n} \mbox{\boldmath $J$} \ket{n}}{Z} \exp(-E_n/k_{\rm B}T). 
\nonumber
\end{equation} 
Here, $g_J$ = 2/7 is the Land\'e g factor, $\mu_{\rm B}$ is the Bohr magneton, and the total angular momentum $J$ = 5/2. 
The eigenvalue $E_n$ and the eigenfunction $\ket{n}$ were obtained by diagonalizing the total Hamiltonian
\begin{equation} 
\begin{split}
H &= H_{\rm_{CEF}} - g_J\mu_{\rm B}\mbox{\boldmath $J$}\cdot(\mu_0\mbox{\boldmath $H$}) \\
&= W\biggl[\frac{x}{F(4)}(O_4^0 + 5O_4^4) + \frac{1-|x|}{F(6)}(O_6^0 - 21O_6^4) \biggr] - g_J\mu_{\rm B}\mbox{\boldmath $J$}\cdot(\mu_0\mbox{\boldmath $H$}).
\end{split}
\nonumber
\end{equation} 
The first term is the CEF Hamiltonian rewritten by the $x$ and $W$ parameters \cite{SM2}, and the second term is the Zeeman term.

\begin{figure}[tb]
\begin{centering}
\includegraphics[width=0.52\textwidth]{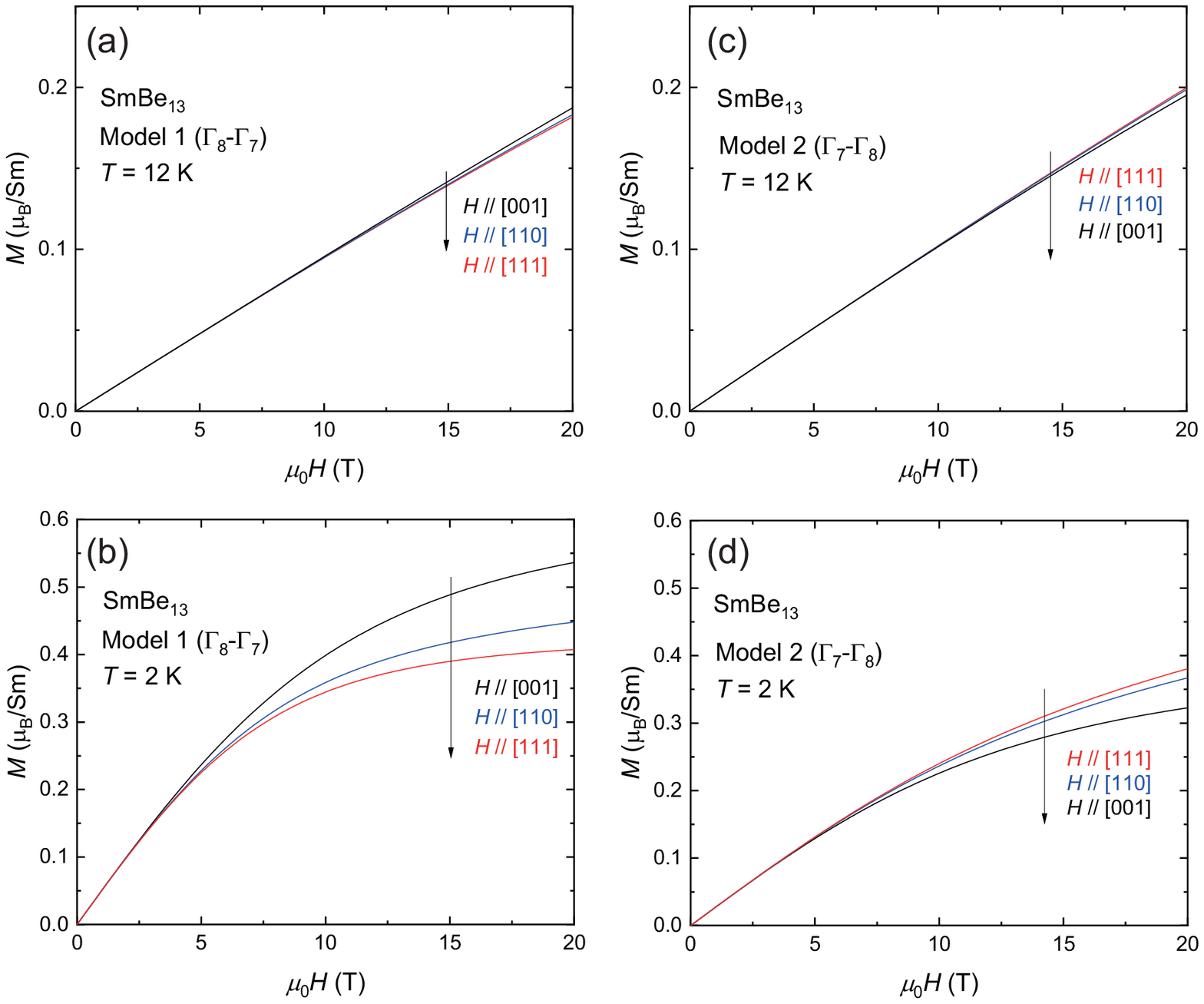}
\caption{
(Color online) The calculated magnetization process based on the CEF effect for model 1 ($\Gamma_8$--$\Gamma_7$ with $\mathit{\Delta}$ = 90 K) at (a) 12 K and (b) 2 K and for model 2 ($\Gamma_7$--$\Gamma_8$ with $\mathit{\Delta}$ = 30 K) at (c) 12 K and (d) 2 K, 
}
\label{FigS4}
\end{centering}
\end{figure}

Figures S4 show the calculated magnetization process for the cubic [001], [011], and [111] directions assuming two CEF level scheme models proposed for SmBe$_{13}$ thus far. 
To investigate the magnetic anisotropy clearly, the calculations were carried out at a lower temperature of 2 K in addition to 12 K, where the effect of the magnetic ordering is excluded. 
Model 1 is the $\Gamma_8$--$\Gamma_7$ CEF level scheme with energy separation $\mathit{\Delta}$ = 90 K ($x$ = 1 and $W$ = -15 K) \cite{Hidaka-SmBe13}, and model 2 is the $\Gamma_7$--$\Gamma_8$ scheme with $\mathit{\Delta}$ = 30 K ($x$ = 1 and $W$ = 5 K). \cite{Besnus}
The magnetization easy axis determined experimentally is consistent with that parallel to [001] for model 1, not parallel to [111] for model 2. 
Thus, in terms of magnetic anisotropy, model 1 provides better description of the CEF state of SmBe$_{13}$. 
Note that the $M$ values for [011] and [111] obtained from the experiments are opposite to the calculated results based on model 1, which might be due to the experimental accuracy. 
The discrepancy in the value of the horizontal axis between the calculated and the experimental results would be improved by taking the molecular field into consideration.

\end{document}